\documentclass[12pt]{article}
\usepackage[a4paper,margin = 2cm]{geometry}
\usepackage{tikz}
\usepackage[OT1]{fontenc}
\usepackage{times}
\usepackage{amsthm}
\usepackage{amssymb}
\usepackage{amsfonts}
\usepackage{setspace}
\usepackage{newtxtext,newtxmath}
\usepackage{graphicx}
\usepackage{rotating}
\usepackage{authblk}
\usepackage{xcolor}
\definecolor{lightblue}{rgb}{0.85,0.00,0.99}

\usepackage{color}
\usepackage[round]{natbib}
\usepackage{bm}
\usepackage{bigints}
\usepackage{multirow}
\usepackage{float}
\usepackage{longtable}
\usepackage{lscape}
\usepackage{setspace}
\usepackage{mathtools}
\usepackage{multicol}
\usepackage{caption, subcaption}
\usepackage{makecell}
\usepackage{pdfpages}
\usepackage[colorlinks,citecolor=blue,urlcolor=blue]{hyperref}

\usetikzlibrary{shapes,decorations,arrows,calc,arrows.meta,fit,arrows,shapes.arrows,
shapes.geometric,shapes.multipart,decorations.pathmorphing,positioning,shapes.swigs}
\tikzset{
    -Latex,auto,node distance =1 cm and 1 cm,semithick,
    state/.style ={ellipse, draw, minimum width = 0.7 cm},
    point/.style = {circle, draw, inner sep=0.04cm,fill,node contents={}},
    bidirected/.style={Latex-Latex,dashed},
    el/.style = {inner sep=2pt, align=left, sloped}
}

\usepackage{bbm}
\newcommand{\hide}[1]{}

\renewcommand{\P}{\mbox{P}}

\newcommand{\E}{\mathrm{E}}
\newcommand\independent{\perp\!\!\!\perp}

\newcommand\dependent{\not\!\perp\!\!\!\perp}

\title{Instrumented difference-in-differences\\ under case-control sampling}
\author[1]{Tran Trong Khoi Le}
\author[1,2]{Emilie Sbidian}
\author[1]{Tat-Thang Vo}
\affil[1]{EPILOGY, Institut Mondor of Biomedical Research, INSERM U955, University Paris Est Créteil, France}
\affil[2]{Department of Dermatology, Henri Mondor Hospital, France}

\begin{document}
\maketitle
\begin{abstract}
Case-control designs are fundamental in epidemiology for the efficient study of rare outcomes. Although instrumental variable (IV) methods have been extended to this setting to address unmeasured confounding, they typically rely on the exclusion restriction assumption, which may be violated when the IV candidates directly affect the outcome through pathways independent of the exposure. In this paper, we propose a novel instrumented difference-in-differences (iDiD) approach tailored to case-control designs. Grounded in structural mean modeling, the proposed method accommodates IV candidates that have time-invariant direct effect on the outcome. When retrospective case-control datasets are collected, the candidate can still be used as a valid instrument on the trend scale when selection bias induced by retrospective sampling is efficiently taken into account. We assess finite-sample performance of this method through extensive simulations, then apply it to evaluate the risk of serious infection of biologic treatments for psoriasis, using French national claim database.
\end{abstract}

\section{Introduction}

In modern epidemiologic research, case–control studies are widely used to investigate rare outcomes and diseases with long induction periods \citep{schlesselman1982case, laugesen2021nordic, ludvigsson2016registers}. Standard case-control analyses typically target conditional treatment effects on the odds ratio scale \citep{cornfield1951method}, obtained via multivariable logistic regression adjusting for treatment and baseline covariates \citep{prentice1979logistic,breslow1980statistical,dey2020practical}. This conventional approach, however, is susceptible to bias from unmeasured or inadequately measured treatment–outcome confounder \citep{rothman2008modern,robins1999choice,ananth2017hidden}, a concern that is particularly important when cases and controls are drawn from administrative or claims databases not originally constructed for research purposes \citep{cadarette2015introduction,fewell2007impact}. 

Instrumental variable (IV) methods are a class of statistical approaches designed to address unmeasured confounding in prospective cohort studies. Recently, these approaches have been extended to case–control settings, with adjustments to account for outcome-dependent sampling that may induce selection bias \citep{bowden2011mendelian}. Standard IV analyses typically rely on an exogeneous variable (i.e. an IV) that is associated with the treatment (i.e. relevance), affects the outcome only through the treatment (i.e. exclusion restriction), and is independent of unmeasured confounders of the treatment–outcome relationship (i.e. independence) \citep{angrist1996identification, baiocchi2014instrumental, hernan2010causal}. The validity of an IV, however, is inherently untestable. For example, when genetic variants are used as instruments, they may influence the outcome through biological pathways that bypass the treatment of interest. This phenomenon, known as horizontal pleiotropy, creates a direct causal pathway between the instrument and the outcome, thereby violating the exclusion restriction \citep{vanderweele2014methodological, sanderson2022mendelian}. Bias may also arise if the unmeasured mediator of the instrument-outcome relationship acts as a confounder of the treatment-outcome relationship, thereby violating independence assumption \citep{martens2006instrumental,swanson2019practical,bowden2011mendelian}. 

To overcome the dual challenges of selection bias from retrospective sampling and the violation of standard IV assumptions, in this paper, we extend the recently proposed instrumented difference-in-differences (iDiD) approach to case-control designs. Our framework, grounded in structural mean modeling (SMM), extends the recent theoretical advancements by \citet{ye2023instrumented} and \citet{vo2024structural}. This method offers two significant advantages over traditional IV approaches. First, by leveraging longitudinal exposure and outcome data at different time points, it accommodates instruments that violate the exclusion restriction through time-invariant direct effects on the outcome. Second, our approach adopts a relaxed independence assumption, which allows for valid treatment effect estimation even when the instrument is associated with unmeasured baseline confounders that has constant effect on the outcome across sampling periods. 

This paper is organized as follows. In Section 2, we provide a technical review of conventional case-control analysis and IV methods for case-control. Section 3 develops the iDiD framework for repeated and cross-sectional case-control sampling, while Section 4 evaluates finite-sample performance of the proposed methods, benchmarking the iDiD estimators against conventional and IV estimators. Finally, we apply the new methods to evaluate the risk of infection associated with biologic therapies for psoriasis, using data from the French National Health Data System.

\section{Case-control and IV for case-control: a brief overview}

Consider a case-control study that evaluates the effect of a binary treatment $D$ on a rare binary outcome $Y$. Cases ($Y=1$) and controls ($Y=0$) are sampled from the subgroup of patients with and without the outcome in the target population, after which data on treatment status $D$ and baseline covariates $X$ are collected in a retrospective manner. Let $Y^d$ denote the outcome value potentially observed under treatment $d\in\{0,1\}$. In a conventional case-control analysis, the target estimand is the conditional treatment effect on the odds ratio scale, which can be formally defined as:
   \[ \theta (X) := \frac{\text{Odds}(Y^{1}=1 \mid X)}{\text{Odds}(Y^{0}=1 \mid X)} \approx \frac{\P(Y^{1}=1 \mid X)}{\P(Y^{0}=1 \mid X)} \] 
Here, the approximation to the conditional risk ratio holds due to the rare outcome condition. 
Under standard causal assumptions, i.e. consistency ($Y^d=Y$ if $D=d$), positivity ($0<\P(D=1\mid X)<1$) and ignorability ($Y^d\independent D\mid X$), an estimator of $\theta(X)$ can be obtained by leveraging a logistic outcome model fitted to the case–control data. For instance, under the following specification:
    \begin{align}
    \text{logit}\{\P(Y=1 \mid D,X)\} = \beta_0 +\beta_1D + \beta_2X,
    \end{align}
the treatment effect $\theta(X) = e^{\beta_1}$ can be estimated by $\hat\theta = e^{\hat\beta_1}$, where $\hat\beta_1$ is an estimator of $\beta_1$. This analysis, however, is biased if there are omitted treatment–outcome confounders. To address unmeasured confounding, one potential approach (adapted from the cohort design literature) is to leverage an exogenous variable $Z$ that satisfies the following structural assumptions:
\begin{itemize}
    \item [(A1)] Relevance: $Z$ is associated with $D$ given $X$, i.e. $D \dependent Z \mid X$.
    \item [(A2)] Independence and exclusion restriction: $Z$ is independent of all unmeasured confounders $U$ that influence the $D \rightarrow Y$ relationship, and only affect $Y$ through its influence on the treatment $D$, i.e., $Z \independent Y^{d} \mid X$ for $d \in \{0,1\}$.
\end{itemize}
The above assumptions constitute the core IV assumptions underlying most instrumental variable approaches. While these assumptions enable the identification of bounds on treatment effects, they are generally insufficient for point identification. Achieving point identification typically requires additional assumptions on treatment effect heterogeneity, which vary across IV methods depending on the target estimand. 
In this work, we focus on the approach of \citet{bowden2011mendelian}, due to its natural extensibility to case–control studies. This approach targets the conditional treatment effect among the treated, defined on the risk ratio scale as:
\[
\beta(X)= \frac{\P(Y^1=1 \mid X,D=1)}{\P(Y^{0}=1 \mid X,D=1)}
\]
To identify $\beta(X)$, the following two assumptions are made:
\begin{itemize}
    \item [(A3)] $\log\beta(X)$ is a finite-dimension function of $X$, e.g. $\log\beta(X)=\beta^\top X$.
    \item[(A4)] The outcome generating mechanism satisfies the following structural mean model (SMM): 
    \begin{align} \label{smm}
    \text{log}\frac{\P(Y=1 \mid D,Z,X)}{\P(Y^{0}=1 \mid D,Z,X)} = \beta(X)\cdot D    
    \end{align}
\end{itemize}
Notably, model (\ref{smm}) implies that the instrument $Z$ does not modify the conditional treatment effect on the risk ratio scale among the exposed. As an illustration, this SMM holds if $(X,U,D,Y)$ follow the causal diagram depicted in Figure \ref{fig:dagA}, and the rare binary outcome is generated according to the log-linear model:
    \[
    \log \P(Y=1\mid D, U,X) = \beta_0+\beta_1D+\beta_2XD+\beta_3X + \beta_4U
    \]
The use of the log link is justified by the rarity of the outcome. In this case, the conditional treatment effect reduces to $\beta(X)= \exp(\beta_1+\beta_2X)$.

Under assumptions (A1)-(A6), \citet{bowden2011mendelian} shows that $\beta(X)$ is identifiable from the observed data via the following moment identity:
    \begin{align} \label{iv}
    \E\big[\{d(Z,X) - \E(d(Z,X)\mid X)\}\cdot Ye^{-D\beta(X)}\big]=0
    \end{align}
where $d(Z,X)$ is an arbitrary function of $(Z,X)$. In the above example, one can consider $d(Z,X) = (Z, X)^\top$, which results in two estimating equations for the two-dimension parameter $(\beta_1,\beta_2)$, assuming one-dimension covariate $X$. Since only the cases contributes information to the expectation on the left-hand side, solving the sample analogue of equation (\ref{iv}) on the case-control data will produce valid estimators for $\beta(X)$, provided that all causal assumptions are satisfied.

\section{Instrumented Difference-in-Differences for case-control} 
\begin{figure}[h]
    \centering
    \begin{subfigure} {0.3\textwidth}
    \centering
        \begin{tikzpicture}[node distance =1.5 cm and 1 cm]
        \node[state,draw = none] (z) at (0,0) {$Z$};
        \node[state,draw = none] (d) [right =of z] {$D$};
        \node[state,draw = none] (y) [right =of d] {$Y$};
        \node[state,circle] (u) at ($(d)!0.5!(y) + (0,-1.2)$) {$U$};
        \node[state, draw = none] (x) at ($(d)!0.5!(y) + (0,1.2)$) {$X$};
        \path (z) edge (d);\path (d) edge (y);
        \path (u) edge (y);\path (u) edge (d); 
        \path (x) edge (d);\path (x) edge (y);
        \end{tikzpicture}
        \caption{}
        \label{fig:dagA}
    \end{subfigure}
    \begin{subfigure} {0.65\textwidth}
    \centering
        \begin{tikzpicture}[node distance =1.5 cm and 1 cm]
        \node[state,draw = none] (z) at (0,0) {$Z$};
        \node[state,draw = none] (d0) [right =of z] {$D_0$};
        \node[state,draw = none] (y0) [right =of d0] {$Y_0$};
        \node[state,draw = none] (d1) [right =of y0] {$D_1$};
        \node[state,draw = none] (y1) [right =of d1] {$Y_1$};  
        \node[state,circle] (u0) at ($(d0)!0.5!(y0) + (0,-1.2)$) {$U_0$};
        \node[state,circle] (u1) at ($(d1)!0.5!(y1) + (0,-1.2)$) {$U_1$};
        \node[state,circle] (u) [above =1cm of d0] {$U$};
        \path (z) edge (d0);
        \path (d0) edge (y0);
        \path(y0) edge (d1);
        \path (d1) edge (y1);
        \path (u0) edge (y0);\path (u0) edge (d0);
        \path (u1) edge (y1);\path (u1) edge (d1);
        \path (u) edge (y0);\path (u) edge (d0); 
        
        \draw [->] (u) to [out=0, in=130] (y1);
        \draw [->] (u) to [out=0, in=130] (d1);
        \draw [->] (z) to [out=40, in =140] (y0);
        \draw [->] (z) to [out=40, in =150] (y1);
        \draw [->] (z) to [out=70, in =200] (u);
        \end{tikzpicture}
        \caption{}
        \label{fig:dagB}
    \end{subfigure}

    \begin{subfigure} {1\textwidth}
    \centering
        \begin{tikzpicture}[node distance =1.5 cm and 1 cm]
        \node[state,draw = none] (z) at (0,0) {$Z$};
        \node[state,draw = none] (d0) [right =of z] {$D_0$};
        \node[state,draw = none] (y0) [right =of d0] {$Y_0$};
        \node[state,draw = none] (d1) [right =of y0] {$D_1$};
        \node[state,draw = none] (y1) [right =of d1] {$Y_1$};  
        \node[state,circle] (u0) at ($(d0)!0.5!(y0) + (0,1.2)$) {$U_0$};
        \node[state,circle] (u1) at ($(d1)!0.5!(y1) + (0,1.2)$) {$U_1$};
        \node[state,draw = none] (x) at ($(d0)!0.5!(y0) + (0,-1.2)$) {$X$};
        \node[state,draw = none] (t) [below =of z] {$T$};
        \node[state,draw = none] (d) [right =of t, below =of d0] {$D$};
        \node[state,draw = none] (y) [right =of d, below =of d1] {$Y$};
        \path (z) edge (d0); \path (z) edge (t);
        \path (d0) edge (y0); \path (d0) edge (d);
        \path(y0) edge (d1); 
        \path (d1) edge (y1);
        \path (u0) edge (y0);\path (u0) edge (d0);
        \path (u1) edge (y1);\path (u1) edge (d1);      
        \path (x) edge (d0); \path (x) edge (y0);
        \path (t) edge (d);

        \draw [->] (x) to [out=180, in =-30] (z);
        \draw [->] (x) to [out=180, in =30] (t);
        \draw [->] (x) to [out=0, in =210] (d1);
        \draw [->] (x) to [out=0, in =210] (y1);
        \draw [->] (d1) to [out=230, in =0] (d);
        \draw [->] (y0) to [out=-70, in =180] (y);
        \draw [->] (y1) to [out=250, in =0] (y);
        \end{tikzpicture}
        \caption{}
        \label{fig:dagC}
    \end{subfigure}
    \caption{Directed acyclic graphs. $(a)$: standard instrumental variable; $(b)$: instrumented difference-in-difference with longitudinal data structure; $(c)$: instrumented difference-in-difference with repeated cross-sectional data structure. In figure 1b, we omit $X$ to simplify the graph.}
    \label{fig:dag}
\end{figure}
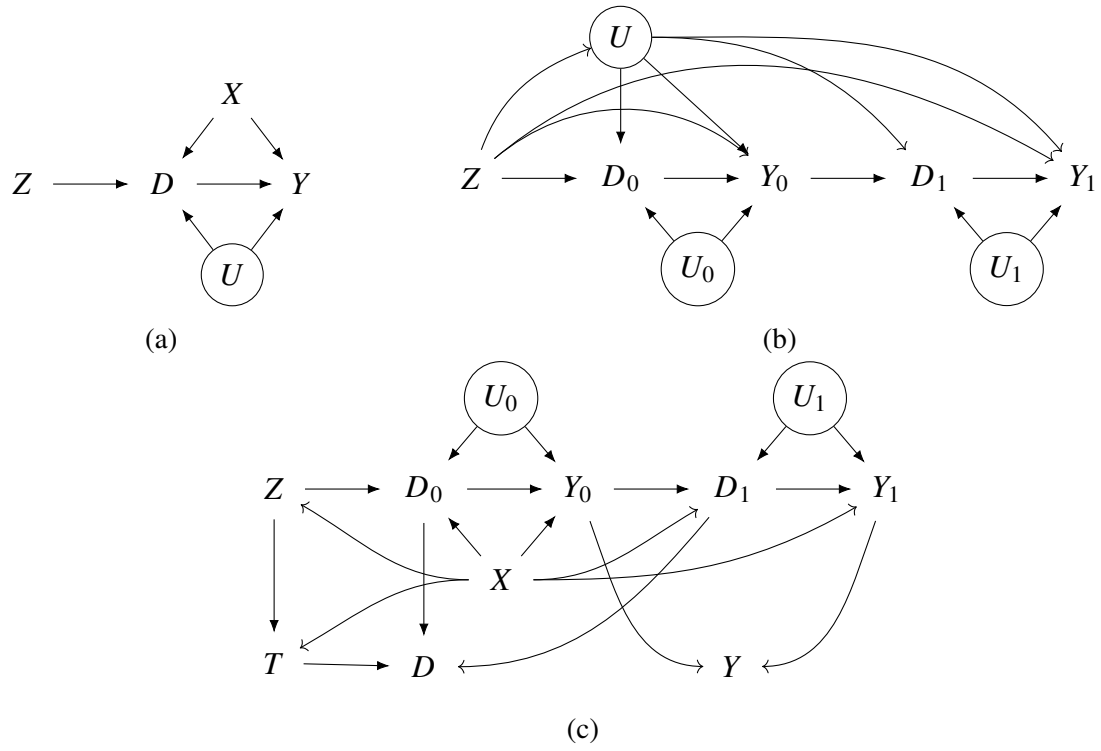

We now proceed to settings where the IV assumptions are violated. For instance, in the causal diagram of Figure \ref{fig:dagB}, $Z$ is not a valid instrument due to a direct effect on $Y$ not mediated by $D$ (violating A3) and its association with the unmeasured confounder $U$ (violating A4). However, if the effects of $Z$ and $U$ on $Y$ are time-invariant, they do not contribute to changes in the outcome over time (the outcome trend), except through their influence on temporal variation in treatment use (the exposure trend). This suggests that $Z$ may remain a valid instrument on the trend scale, motivating the so-called instrumented difference-in-differences (IDiD) methods suggested by \citet{ye2023instrumented} and by \citet{vo2024structural}.

To formalize IDiD, suppose that the target population is followed over two periods, i.e. $t=0$ and $t = 1$. Let $(D_t,Y_t)$ denote the exposure and outcome status at period $t$, $U$ baseline unmeasured confounders and  $U_t$ time-varying unmeasured confounders at period $t$. We allow for a general dependence structure between the instrument $Z$ and the variables $U,U_t,D_t$ and $Y_t$, as depicted in Figure \ref{fig:dagB}.
Let $Y^{d}_t$ denote the counterfactual outcome at time $t$ under treatment $D_t=d \in \{0,1\}$. Following \citet{bowden2011mendelian}, we focus on the conditional and marginal average treatment
effect among the treated (ATT), assuming these are time-invariant and that there are no carry-over effects across time points, i.e.:
    \[\beta(X) = \frac{\P(Y^{1}_t=1\mid D_t=1,X)}{\P(Y^{0}_t=1\mid D_t=1,X)} \qquad \text{and} \qquad
    ATT=\frac{\P(Y^{1}_t=1\mid D_t=1)}{\P(Y^{0}_t=1\mid D_t=1)}\]
To identify $\beta(X)$ and $ATT$, \citet{vo2024structural} impose assumptions (A3), along with:
\begin{itemize}
    \item [(A5)] No $T-Z$ interaction: $Z$ does not modify the effect of time on the outcome, in the sense that $\P(Y_1^{d}=1\mid X, Z) = \P(Y_0^{d}=1\mid X, Z)e^{ m(X,\gamma)}$, where $d\in \{0,1\}$ and $m(x,\gamma)$ is a pre-specified parametric function of $x$.
    \item[(A6)] At each time point $t$, the outcome generating mechanism satisfies model (\ref{smm}), i.e., $Z$ does not modify the effect of the treatment on the multiplicative scale among the exposed: 
    \begin{align} \label{smm.idid}
    \text{log}\frac{\P(Y_t=1 \mid D_t,Z,X)}{\P(Y_t^{0}=1 \mid D_t,Z,X)} = \beta(X)\cdot D_t    
    \end{align}
\end{itemize}
Intuitively, the absence of time and treatment effect modification by $Z$ reflect the independence and restriction assumptions on the trend scale that $Z$ needs to satisfy (see \citet{ye2023instrumented} and \citet{vo2024structural} for formal explanations). To illustrate the utility of these two assumptions, consider the following log-linear outcome models:
\begin{align*}
    \P(Y_1=1|\overline{D}_1, \overline{U}_1, X, Z)&= \exp(\beta_0^1+\fbox{$\beta_1D_1+\beta_2D_1X$}+\beta_3U_1+\fbox{$\beta_4U+\beta_5Z$} + \beta_6^1X)\\
    \P(Y_0=1|D_0, \overline{U}_0, X, Z) &= \exp(\beta_0^0+\fbox{$\beta_1D_0+\beta_2D_0X$}+\beta_3U_0+\fbox{$\beta_4U+\beta_5Z$} + \beta_6^0X)
\end{align*}
Here, $U_0$ and $U_1$ are equal in distribution given $X$, and the boxed terms highlight time-invariant components of the outcome model. Notably, these outcome models satisfy the SMM (\ref{smm}) with $\beta(X)=\beta_1+ \beta_2X$ (see the Online Supplementary Material for a formal proof). However, the presence of $Z$ in these models, together with its association with the unmeasured confounder $U$ not fully explained by $X$ (i.e. $U\dependent Z\mid X$), precludes the use of $Z$ as a valid (standard) IV. When only cross-sectional data at one time point are available, naively applying standard IV methods would therefore yield biased estimates of the treatment effect. In contrast, assumption (A7) and (A8) holds in this setting due to the time-invariant effect of $Z$ and $U$ on the outcome, with $m(X,\gamma)=\gamma_0+\gamma_1X$ where $\gamma_0=\beta_0^1-\beta_0^0$ and $\gamma_1=\beta_6^1-\beta_6^0$. This enables the identification of 
$\beta(X)$ from the observed (longitudinal) data through the moment identity:
    \begin{align}\label{idid}
        E\big\{f(X,Z)\cdot [Y_1 e^{-\beta(X) D_1} - Y_0 e^{-\beta(X)D_0 + m(X,\gamma)}]\big\} = 0 
    \end{align}
where $f(X,Z)$ is an arbitrary function of $(X,Z)$. For instance, one can consider $f(X,Z)=(1,X,Z,XZ)$, yielding four estimating functions for the four-dimension parameter $(\beta_1,\beta_2,\gamma_0,\gamma_1)$ when $X$ is one-dimensional. Applying these estimating equations on case-control data, however, is non-trivial, as they require information across two timepoints. To tackle this challenge, in the rest of this section, we propose two case-control designs that enables the estimation of the proposed treatment effects based on equation (4). The first design assumes that  a case-control sample is available at each time point $t=0,1$. The second design relaxes this requirement by considering a single case–control study that collects data from a repeated cross-sectional cohort, assuming that the timing of recruitment is independent of treatment and outcome status conditional on $Z$ and $X$.  


\subsection{Repeated case-control sampling} 
Suppose that two case-control studies are independently conducted in the same target population at $t=0$ and $t=1$. In the study conducted at time $t$, a representative sample of cases (i.e. those with $Y_t = 1$), and of controls (i.e. those with $Y_t = 0$) is selected. Let $S_t$ be a binary sampling indicator, such that $S_t=1$ if an individual is recruited into the study conducted at time $t$. Our proposal accommodates potential overlap between the two studies, that is, the existence of individuals for whom $S_0 = S_1 = 1$ thereby encompassing both studies. Of note, the proportion of cases $\pi_{S=t}$ in the sample (or the sampling fraction) at time $t$, and the sample size $n_t$ at time $t$ are fixed \textit{a priori} by study design. 
A comparable setting is when a single case–control study is conducted with recruitment occurring at two distinct time points, $t=0,1$. In this case, the two-phase recruitment is prespecified in the study protocol, with $\pi_{S=t}$ and $n_t$ determined in advance. This setting differs from a cross-sectional case–control study with a single recruitment phase, which is considered in the next subsection.

Let $\pi_t := \P(Y_t = 1)$ denote the marginal probability of the outcome at time $t \in \{0,1\}$. We assume that $\pi_t$ is either known \textit{a priori} or can be consistently estimated from an independent population-based study. Alternatively, knowledge of the relative change in outcome prevalence over time, i.e., $\pi_1 / \pi_0$, is sufficient. This requirement for external information on disease prevalence is common in the literature on causal inference with retrospective data (e.g., \citet{rose2008simple}), as it provides the necessary scaling to recover population-level causal effects from sample-based distributions.


Specifically, a naive application of equation (\ref{idid}) on the two case-control studies will lead to biased estimate of $\beta(X)$, due to the oversampling of the cases at each time point. To correct for this, we leverage the fact that while the marginal distribution of $Y_t$ is fixed by design, the case-control samples provide unbiased information on the distribution of covariates conditional on event status. More precisely, \[\P(X,Z,D_t\mid Y_t=y) = \P^*(X,Z,D_t\mid S_t = 1, Y_t = y)\] 
for $t,y \in \{0,1\}$, where $\P^*$ is the sampling distribution underlying the two case-control studies. As formally demonstrated in the Online Supplementary Material, incorporating appropriate sampling weights into the moment condition (\ref{idid}) yields the following estimating equation for $(\beta,\gamma)$:
\begin{align}
    \frac{1}{n_0}\frac{\pi_0}{\pi_{S=0}}\sum_{i=1}^{n}f(X_i,Z_i)S_{0i}Y_{0i}e^{-\beta(X_i)D_{0i} + m(X_i,\gamma)} = \frac{1}{n_1}\frac{\pi_1}{\pi_{S=1}}\sum_{i=1}^{n}f(X_i,Z_i)S_{1i}Y_{1i}e^{-\beta(X_i)D_{1i}}  
\end{align}
This leads to a time-specific estimator $\widehat{ATT}_t$ for the marginal risk ratio $ATT$:
\[
\widehat{ATT}_t=\frac{\sum_{i:S_{ti}=1}Y_{ti}D_{ti}}{\sum_{i:S_{ti}=1}Y_{ti}D_{ti}\hat\beta^{-1}(X_i)}
\]
where $\hat\beta(X)$ is obtained by solving equation (6). A final estimator of $ATT$ can then be constructed by averaging the two time-specific estimates.

By standard M-estimation theory, $(\hat\beta, \hat\gamma, \widehat{ATT}_t)$ is asymptotically normal. The variance of these estimators can be consistently estimated using the sandwich estimator, i.e. 
$\widehat{\text{Var}}(\hat{\beta}, \hat{\gamma},\widehat{ATT}_t) = \frac{1}{n} A_n^{-1} B_n (A_n^{-1})^\top$, 
where $A_n$ and $B_n$ denote the empirical Jacobian matrix and the empirical covariance of the estimating function, respectively, defined under the sampling distribution $\P^*$ as 
\[A_n = \E^* \left[ \frac{\partial \psi(\beta, \gamma,ATT)}{\partial \beta, \gamma,ATT} \right], \quad B_n = \E^* \left[ \psi(\beta, \gamma, ATT)\cdot \psi(\beta, \gamma, ATT)^\top \right] \]
where the estimating function for $(\beta,\gamma)$ is given by:
\[
\psi_{\beta,\gamma} = f(X,Z)S_1Y_1e^{-\beta(X)D_1}\frac{\pi_1}{p_1\pi_{S=1}} - f(X,Z)S_0Y_0e^{-\beta(X)D_0 + m(X,\gamma)}\frac{\pi_0}{p_0\pi_{S=0}}
\]
and the estimating function for $ATT$, based on data from time $t$ is $\psi_{ATT_t} = D_tY_t - ATT\cdot D_tY_t\beta^{-1}(X)$. Here, $p_t=n_t/(n_0+n_1)$ represents the sampling fraction of the target population at time $t \in \{0,1\}$. 

The proposed approach accommodates any choice of function $f(X,Z)$. In practice, however, it is preferable to use functions involving no or only simple nuisance parameters. For example, analysts may wish to center the covariates by defining $X^* = X - \E(X)$ in the analysis to facilitate interpretation of the conditional treatment effect. In that case, $\E(X)$ can be estimated from the case-control data at time $t$ as:
$\hat \E_t(X) = \sum_{i:S_i=t} W_i X_i(\sum_{i:S_i=t} W_i)^{-1}$, 
where the weights are given by $W_i = Y_{it}\pi_t/p_t + (1-Y_{it})(1-\pi_t)/p_t$. Alternatively, when the event prevalence $\pi_t$ is small, $\E(X)$ can be approximated by the sample mean of the controls at time $t$, as controls provide a close approximation to the target population. In both cases, the estimating function for $\E(X)$ needs to be added into $\psi(\beta,\gamma,ATT)$ to account for the additional variability induced by estimating $\E(X)$.

Finally, the assumption that the event prevalence $\pi_t$ is known \textit{a priori} can be relaxed when additional data from the target population are available. For instance, this arises when the case-control studies are nested within a cohort that is representative of the target population, allowing $\pi_t$ to be estimated from the source cohort. Alternatively, one may consider a case--cohort design comprising (i) a random sample of cases from the target population at time $t$, and (ii) a random subcohort drawn at baseline and followed over time to observe event occurrence. The latter provides a basis for estimating $\pi_t$. In both settings (i.e., nested case--control and case--cohort designs), the resulting estimate $\hat\pi_t$ can be substituted into equation~(5) to estimate $(\beta,\gamma)$. However, additional adjustments are required to account for the variability introduced by estimating $\pi_t$. 

\subsection{Cross-sectional case-control sampling}
Suppose now that the target population follows a repeated cross-sectional (or “pseudo-longitudinal”) data structure, in which the treatment and outcome status of each individual are observed at a single time point. Specifically, we observe $O=(D,Y,X,Z,T)$ where $T=0,1$ indicates the time period, $D=D_1I(T = 1) + D_0I(T=0)$ and $Y = Y_1I(T = 1) + Y_0I(T=0)$. This setup arises naturally in many applied settings, including population-based surveys and administrative data, where temporal variation is captured through independent samples rather than longitudinal follow-up of individuals. 

Importantly, this data structure renders standard IV methods invalid even when assumptions (A1)–(A2) hold at each time point. The bias arises from ignoring selection into time periods. As illustrated in Figure \ref{fig:dagC}, the observed outcome $Y$ is a mixture of $Y_0$ and $Y_1$, which induces post-treatment pathways such as 
$Z \rightarrow D_0 \rightarrow Y_0 \rightarrow Y$ and 
$Z \rightarrow D_0 \rightarrow Y_0 \rightarrow D_1 \rightarrow Y_1 \rightarrow Y$. 
These pathways create direct associations between $Z$ and $Y$ that do not operate exclusively through the observed treatment status $D$, thereby violating the exclusion restriction and invalidating IV analyses based solely on $(Z, X, D, Y)$. An exception arises under the sharp null of no individual-level treatment effect at either time point, which blocks these pathways; however, this assumption is difficult (even impossible) to verify in the presence of unmeasured confounders $(U_0,U_1)$.

To account for the fact that individuals observed at time $t$ may not be representative of the target population, we impose the following assumption on the observation mechanism: 
\begin{itemize}
    \item [(A7)] Selection into each time period is as good as random given $X$ and $Z$, i.e. $T\independent (D_0,Y_0,D_1,Y_1)\mid X,Z$.
\end{itemize}
This assumption ensures that, conditional on $(X,Z)$, the composition of individuals observed at each time point is comparable, so that differences across periods reflect population-level changes rather than selection effects. Assumption (A9) may be violated in several practical settings. For instance, individuals may be more likely to be included in the data if they experience the outcome (e.g., hospitalizations captured in administrative databases) or if treated individuals being more closely monitored. In such cases, selection into time periods induces additional bias that cannot be removed by conditioning on $(X,Z)$ alone.

Under assumption (A1)--(A4) and (A7), $\beta(X)$ is identifiable from the observed data through a weighting-based moment condition that reweights observations by the inverse probability of being observed at each time point. For example, under standard IV assumptions (A1)–(A4), $\beta(X)$ satisfies:
\begin{align} \label{iv1}
    \E\bigg[\frac{\{d(Z,X) - \E(d(Z,X)\mid X)\}\cdot Ye^{-D\beta(X)}I(T=t)}{\P(T=t\mid X,Z)}\bigg]=0
\end{align}
In contrast, under iDiD assumptions (A3) and (A5)–(A7), $\beta(X)$ satisfies:
\begin{align} \label{idid1}
    \E\bigg\{f(X,Z)\times \bigg[\frac{TYe^{-\beta(X) D}}{\P(T = 1|Z,X)} - \frac{(1 - T)Ye^{-\beta( X) D+ m( X)}}{\P(T = 0\mid Z, X)}\bigg]\bigg\} = 0
\end{align}
Notably, constructing estimators based on these moment conditions does not require knowledge of the marginal outcome prevalence $\P(Y=1)$, but does require estimation of the selection probability $\P(T=1 \mid X,Z)$. To this end, we assume that $\P(T=1 \mid X,Z)$ follows the logistic regression model:
\begin{align}\label{est.t.cross}
\P(T=1\mid X,Z) = g(X,Z,\eta)=\mathrm{expit}(\eta_0 + \eta_1X + \eta_2Z)
\end{align}
Now consider a case–control study in which representative samples of cases (i.e., individuals with $Y=1$) and controls (i.e., individuals with $Y=0$) are drawn from the target population. Under the rare outcome assumption, the control group provides a reasonable approximation to the covariate distribution in the target population, which in turn enables consistent (or approximately consistent) estimation of the model for $\P(T=1\mid X,Z)$.
Motivated by this observation, we propose the following IDiD estimation procedure for $\beta(X)$ and $ATT$ based on case-control data. 
\begin{itemize}
    \item [S1.] Estimate $\eta$ by fitting model \eqref{est.t.cross} on the control data $Y=0$ via maximum likelihood.
    \item [S2.] Obtain an estimate $(\hat\beta,\hat\gamma)$ for $(\beta,\gamma)$ by solving the sample analog of the moment (\ref{idid1}) on the case data:
    \[\sum_{i:Y_i=1} f(X_i,Z_i)\times \bigg[\frac{T_ie^{-\beta(X_i) D_i}}{g(X_i,Z_i,\hat\eta)} - \frac{(1 - T_i)e^{-\beta( X_i) D_i+ m( X_i,\gamma)}}{1-g(X_i,Z_i,\hat\eta)}\bigg] = 0\]
    where $\hat\eta$ is an estimate of $\eta$ obtained from step 1.
    \item [S3.] A time-specific estimator $\widehat{ATT}_t$ for $ATT$ can be expressed as:
    \[
    \widehat{ATT}_t=\frac{\sum_{i:Y_i=1}DI(T=t)\cdot \hat\P(T=t\mid X,Z)}{\sum_{i:Y_i=1}DI(T=t)\cdot \hat\beta^{-1}(X)\cdot \hat\P(T=t\mid X,Z)}
    \]
    where $\hat\P(T=t\mid X,Z)$ is an estimate of $\P(T=t\mid X,Z)$.  A final estimator of $ATT$ can then be constructed by averaging the two time-specific estimates.
\end{itemize}
An analogous IV-based estimation procedure for $\beta(X)$ and $ATT$ can also be constructed from equation~\eqref{iv1} (details are provided in the Online Supplementary Materials).

By M-estimation theory, the joint estimator $(\hat\beta,\hat\gamma,\hat\eta,\widehat{ATT}_t)$ is asymptotically normal. Its covariance matrix can be estimated by the sandwich estimator $\widehat{\text{Var}}(\hat{\beta}, \hat{\gamma}, \hat\eta,\widehat{ATT}_t) = \frac{1}{n} A_n^{-1} B_n (A_n^{-1})^\top$, where $A_n$ and $B_n$ denote the empirical Jacobian matrix and the empirical covariance of the estimating function $\psi$ of $(\beta,\gamma,\eta,ATT)$, under the sampling distribution $\P^*$. Specifically, the estimating function for $(\beta,\gamma)$ is:
\begin{align*}
    \psi_{\beta,\gamma} &= f(X,Z)\frac{TYe^{-\beta(X)D}}{g(X,Z,\eta)} - f(X,Z)\frac{(1-T)Ye^{-\beta(X)D + m(X,\gamma)}}{1-g(X,Z,\eta)}
\end{align*}
and the estimating function for ATT, based on data from time $t$ is 
\[\psi_{ATT_t}=DYI(T=t)\cdot \P(T=t\mid X,Z)-ATT\cdot DYI(T=t)\beta^{-1}(X) \cdot \P(T=t\mid X,Z).\]
\section{Simulation study}
In this section, we conduct a series of simulation studies to evaluate the finite-sample performance of the proposed iDiD approach in comparison with standard IV.
\paragraph{Scenario 1.}
We consider a longitudinal data-generating process in which the outcome $Y_t$ follows a log-linear model that includes $D_t$, $Z$, $X$ and $U_t$, i.e.:
\begin{align*}
    X&=\text{min}\{\text{Poisson}(0.5) + 0.5, 3.5\}\\
    U_t &= \text{Unif}(0,2)\\
    \P(Z=1\mid X) &= \operatorname{expit}(-0.8 + X)\\
    \P(D_t=1\mid U_t,X,Z) &= \text{expit}(\delta_{0,t} + \delta_{U_t,t}U_t + \delta_{Z,t}Z + \delta_{X,t}X + Y_0 \mathbf{1}_{t=1})\\
    \P(Y_t = 1 \mid D_t, U_t, X, Z)
    &= \exp\big(\beta_{0,t} - 0.2 D_t + \beta_{U,t} U_t + \beta_{X,t} X + \beta_Z Z\big).
\end{align*}
We specify two values for $\beta_Z$, i.e. $\beta_Z \in \{0, 0.15\}$. The instrument $Z$ satisfies the exclusion restriction required for standard IV methods when $\beta_Z = 0$, but not when $\beta_Z = 0.15$. Other parameters in the data generating models are set up such that the event prevalence is approximately $1\%$, $5\%$, $10\%$ and $20\%$ at each time point, and other standard IV assumptions are satisfied (see more details in the Online Supplementary Materials).

At each time point, an independent case--control sample is drawn from the target population, consisting of $n$ cases and $5n$ controls, where $n \in \{250, 500, 1000, 2000, 3000\}$. Three approaches are then considered for estimating the treatment effect $\beta(X)=ATT=\exp(-0.2)$:
\begin{itemize}
    \item[(i)] \textbf{Conventional logistic regression}: A standard logistic regression model is fitted to the case-control sample at each time point, adjusting for treatment status at such time point and observed covariates.
    
    \item[(ii)] \textbf{SMM for standard IV}: The SMM method proposed by \citet{bowden2011mendelian} is applied separately on the case-control data at each time point, with correctly specified SMM.
    
    \item[(iii)] \textbf{SMM for iDiD}: The proposed iDiD method is applied to the case-control samples across time points, with correctly specified SMM.
\end{itemize}

\paragraph{Scenario 2.}
We next consider a repeated cross-sectional data structure. Specifically, after generating $(D_t, Y_t)$ for $t = 0,1$ as in Scenario 1, we generate selection into each time period according to the logistic model
\[
P(T = 1 \mid X, Z)
= \mathrm{expit}(0.12 + 0.23 Z - 0.98 X).
\]
A single case--control sample is then drawn from the target population, consisting of $n$ cases and $5n$ controls, where $n \in \{250, 500, 1000, 2000, 3000\}$. Three approaches analoguous to those in Scenario 1 are then considered for estimating the treatment effect, namely (i) Conventional logistic regression, (ii) SMM for standard IV (ignoring the temporal aspect) and (iii) SMM for iDiD (with the SMM and logistic selection models both correctly specified). 

\paragraph{Bias and coverage evaluation.} Across both scenarios, we evaluate (i) the absolute bias of the resulting estimator $\hat{\beta}_D$ relative to the true parameter value $\beta_D=-0.2$, and (ii) the empirical coverage probability of the corresponding 95\% Wald confidence interval for $\beta_D$. For each setting, results are based on 5000 iterations.

\paragraph{Simulation results -- Scenario 1 (longitudinal data structure).}
The conventional logistic regression approach exhibits substantial bias and poor coverage due to unmeasured confounding. 

In contrast, both the standard IV and iDiD estimators yield consistent treatment effect estimates with appropriate 95\% CI coverage when $Z$ is a valid instrument (i.e., $\beta_Z = 0$). Notably, however, the standard IV estimator exhibits appreciable finite-sample bias and suboptimal coverage at smaller sample sizes compared with the iDiD estimator (Figure \ref{fig:valid_ncross}).

When $Z$ is an invalid instrument (i.e., $\beta_Z = 0.15$), the standard IV method applied separately to cross-sectional data at each time point becomes severely biased, leading to poor 95\% CI coverage (Figure \ref{fig:invalid_ncross}). In contrast, the iDiD estimator continues to provide consistent treatment effect estimates with valid 95\% CI coverage.

\paragraph{Simulation results -- Scenario 2 (cross-sectional data structure).}
Even when the IV assumptions hold (i.e., $\beta_Z = 0$), the standard IV estimator exhibits substantial bias and undercoverage of the corresponding 95\%CI, as it ignores the temporal structure of the data (Figure \ref{fig:valid_cross}).

In contrast, the proposed iDiD method remains robust, yielding consistent treatment effect estimates with valid 95\%CI coverage when the outcome is sufficiently rare (e.g., $\leq 10\%$). As the event prevalence increases, however, the control group becomes less representative of the target population. This, in turn, induces bias in the estimation of the selection model $P(T = 1 \mid X, Z)$ when it is fitted using control data, ultimately leading to biased treatment effect estimates (Figure \ref{fig:invalid_cross}).

\section{Application on real data}
We now apply the proposed method to evaluate the effect of modern biologic therapies (\(D=1\); interleukin-17 inhibitors and interleukin-23 inhibitors) versus conventional biologic therapies (\(D=0\); tumor necrosis factor inhibitors and interleukin-12/23 inhibitors) on the risk of serious infection among patients with psoriasis, using claims data from the French National Health Data System (Syst\`eme National des Donn\'ees de Sant\'e [SNDS]).

Previous studies using SNDS data have established a cohort of psoriatic patients who initiated biologic treatment between January 1, 2016 and December 31, 2022 \citep{le2026infection}. In this cohort, patients were followed until the occurrence of a serious infection leading to hospitalization or until two years after treatment initiation, whichever occurred first.

\begin{figure}
    \centering
    \includegraphics[width=1\linewidth]{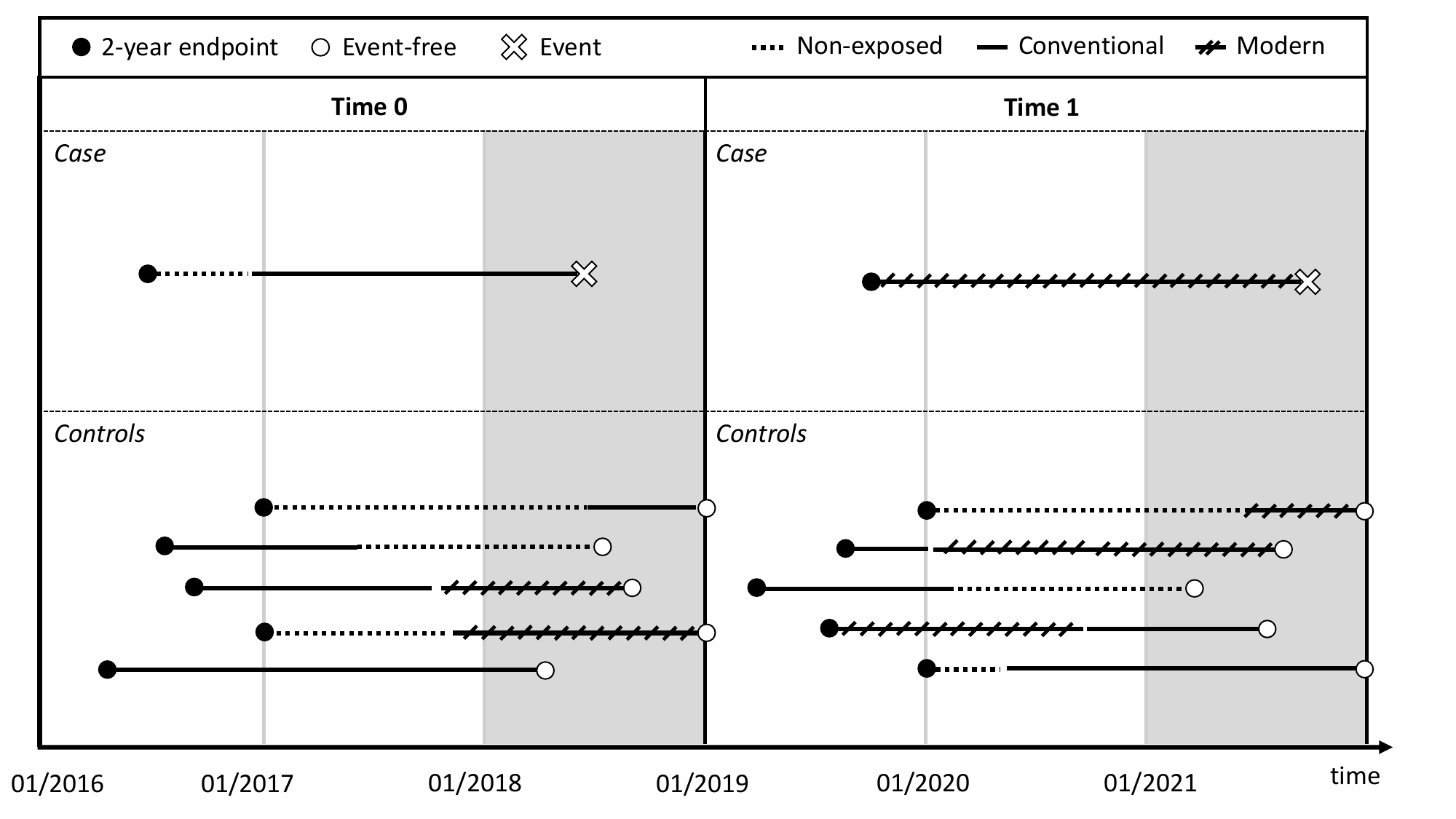}
    \caption{Illstrative data analysis: selection of case and controls.}
    \label{fig:cc_design}
\end{figure}

In this illustrative analysis, we considered two time points, \(t=0\) and \(t=1\), corresponding to calendar years 2018 and 2021, respectively (Figure \ref{fig:cc_design}). These time points were selected to ensure that each patient contribute information to at most one period, thereby yielding a repeated cross-sectional design.

We employed a binary geographical instrument. For each administrative department in France, we first calculated the change in exposure prevalence between \(t=0\) and \(t=1\). Departments were assigned $Z=1$ if this change exceeded the $\tau$-quantile of the overall distribution, and $Z=0$ otherwise. To identify the optimal value for $\tau$, we performed a grid search from $\tau=0.25$ to $\tau=0.75$, selected the value that maximized the correlation between $Z$ and $D$. The optimal value of $\tau$ obtained from this procedure is $0.4$, yielding a ``rapid change'' group (\(Z=1\)), in which the prevalence of modern biologic prescribing increased from \(23\%\) to \(55\%\), and a ``slow change'' group (\(Z=0\)), in which prescribing increased from \(23\%\) to \(39\%\).

Within each period \(t\), all cases (patients experiencing serious infection) were included. Controls were randomly sampled at a 1:5 case--control ratio from individuals who did not experience serious infection during period \(t\). Exposure status for each selected individual was defined according to the biologic therapy to which they were predominantly exposed during follow-up.

Data on six baseline covariates were retrospectively collected for all selected individuals: age, sex, baseline comorbidity burden, baseline medication use, history of serious infection, and baseline anti-infective drug use (Table \ref{tab:1}).

The resulting case--control sample included 2,772 patients (462 cases and 2,310 controls). At each time point \(t\), we assumed that the IDiD assumptions (A7) and (A8) held, with \(m(X,\gamma)=\gamma^\top X\) and \(\beta(X)=\beta\). Additionally, we specified a logistic regression model including all baseline covariates and the instrument linearly to characterize the relationship between period membership \(T\) and \((X,Z)\).

Application of the proposed approach yielded an estimated treatment effect of \(\hat{\beta}=-0.54\) (95\% CI: $-2.04$, $0.95$). This estimate suggests that, among individuals who actually received modern biologics, the risk of serious infection would have been approximately 1.72 times higher had they instead been exposed to conventional biologics. However, the estimated association did not reach statistical significance.

The \texttt{R} code used to implement this analysis is available at \href{https://github.com/lttk/idid}{https://github.com/lttk/idid}.

\begin{center}
    \begin{table}[H]
    \captionsetup{margin=0cm, font={stretch=1}}
    \centering
    \caption{Illustrative data analysis: Description of the cases and controls across two time points}
    \label{tab:1}
    \scalebox{0.75}{
    \renewcommand{\arraystretch}{1.3}
    \begin{tabular}{lrrrrrr}
    \hline
    Characteristics & \multicolumn{2}{c}{Time 0} & \multicolumn{2}{c}{Time 1} & \multicolumn{2}{c}{Overall} \\
    \hline
     & Cases & Controls & Cases & Controls & Cases & Controls \\
    No. & 186 & 977 & 276 & 1333 & 462 & 2310 \\
    Age (mean, SD) & 54.57 (15.45) & 48.15 (14.14) & 56.84 (15.75) & 48.13 (14.31) & 55.93 (15.65) & 48.14 (14.24)\\
    Sex (no., \%) & 84 (45.16) & 422 (43.19) & 121 (43.84) & 614 (46.06) & 205 (44.37) & 1036 (44.85)\\
    Infectious hospitalisation (no., \%) & 36 (19.35) & 64 (6.55) & 52 (18.84) & 68 (5.10) & 88 (19.05) & 132 (5.71)\\
    Comorbidity (no., \%) & 48 (25.81) & 258 (26.41) & 63 (22.83) & 341 (25.58) & 111 (24.03) & 599 (25.93)\\
    Anti-infective (no., \%) & 41 (22.04) & 263 (26.92) & 78 (28.26) & 349 (26.18) & 119 (25.76) & 612 (26.49)\\
    Medication (no., \%) & 81 (43.55) & 475 (48.62) & 124 (44.93) & 625 (46.89) & 205 (44.37) & 1100 (47.62)\\
    Modern treatment (no., \%) & 43 (23.12) & 235 (24.05) & 111 (40.22) & 644 (48.31) & 154 (33.33) & 879 (38.05)\\
    Instrument = 1 (no., \%) & 120 (64.52) & 618 (63.25) & 164 (59.42) & 833 (62.49) & 284 (61.47) & 1451 (62.81)\\
\hline
    \end{tabular}}
    \end{table}
\end{center}

\section{Discussion}

In this paper, we proposed a novel instrumented difference-in-differences approach tailored to case-control designs. Grounded in structural mean modeling, the proposed method accommodate outcome-dependent sampling in settings where standard instrumental variable assumptions of independence or exclusion restriction may be violated. By leveraging information across multiple time points, the new method removes the direct, time-invariant effect of the instrument (or unmeasured confounders associated with the instrument) on the outcome, thereby relaxing traditional IV requirements.
The flexibility of this so-called IDiD framework significantly expands the selection of viable instruments in practice, where common instruments such as geographic regions, healthcare policy jurisdictions, insurance types, physician preference and genetic variants are frequently critiqued due to potential pleiotropy or direct effects on the outcome that bypass the exposure \citep{baiocchi2014instrumental}. When the direct relationship of these instrument candidates with the outcome remains stable across study periods, they are more suitable to be used as instruments on the trend scale, as suggested by our proposed method. 

Several directions for future research remain. First, the current development adopts many parametric assumptions, e.g., a logistic regression model for the period membership in repeated cross-sectional settings. Extending the framework to allow more flexible modeling assumption may improve robustness to model misspecification. Second, future research may extend the methodology to accommodate more complex retrospective designs such as nested case-control sampling or standard case-control with continuous exposures. Finally, additional methodological work is needed to develop diagnostic tools and falsification procedures for assessing the plausibility of the identifying assumptions in applied settings, particularly when instruments are derived in an ad-hoc manner from geographical or policy variation as in our illustrative data analysis.

\section{Funding}
T.T.V is supported by the French National Research Agency (Agence Nationale de la Recherche), through a funding for Chaires de Professeur Junior (23R09551S-MEDIATION).
\section{Conflict of interest}
All authors declare that they have no conflicts of interest.
\section{Data availability}
The data underlying this article were provided by French data protection agency (\textit{Commission nationale de l’informatique et des libertés} [CNIL]: MLD/TDC/AR213829). Informed consent was not required to use pseudonymized data in SNDS.

\bibliographystyle{plainnat}
\bibliography{reference}
\newpage

\section*{Tables and Figures}
 
\begin{landscape}
\begin{figure}[H]
    \centering
    \begin{subfigure} {0.75\textwidth}
    \centering
      \centering
      \includegraphics[width=1\textwidth]{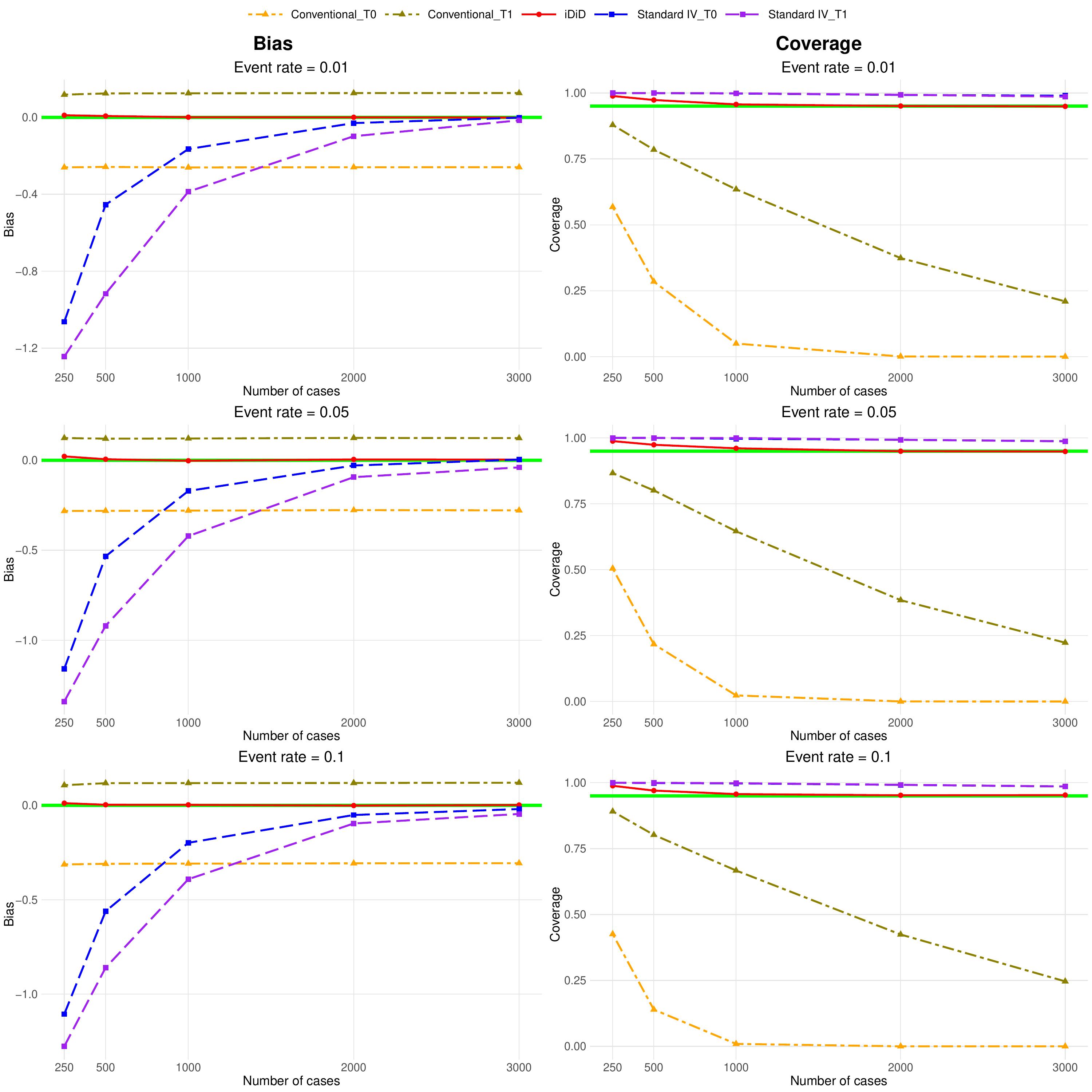}
      \caption{Valid standard IV}
      \label{fig:valid_ncross}
    \end{subfigure}
    \begin{subfigure} {0.75\textwidth}
      \centering
      \includegraphics[width=1\textwidth]{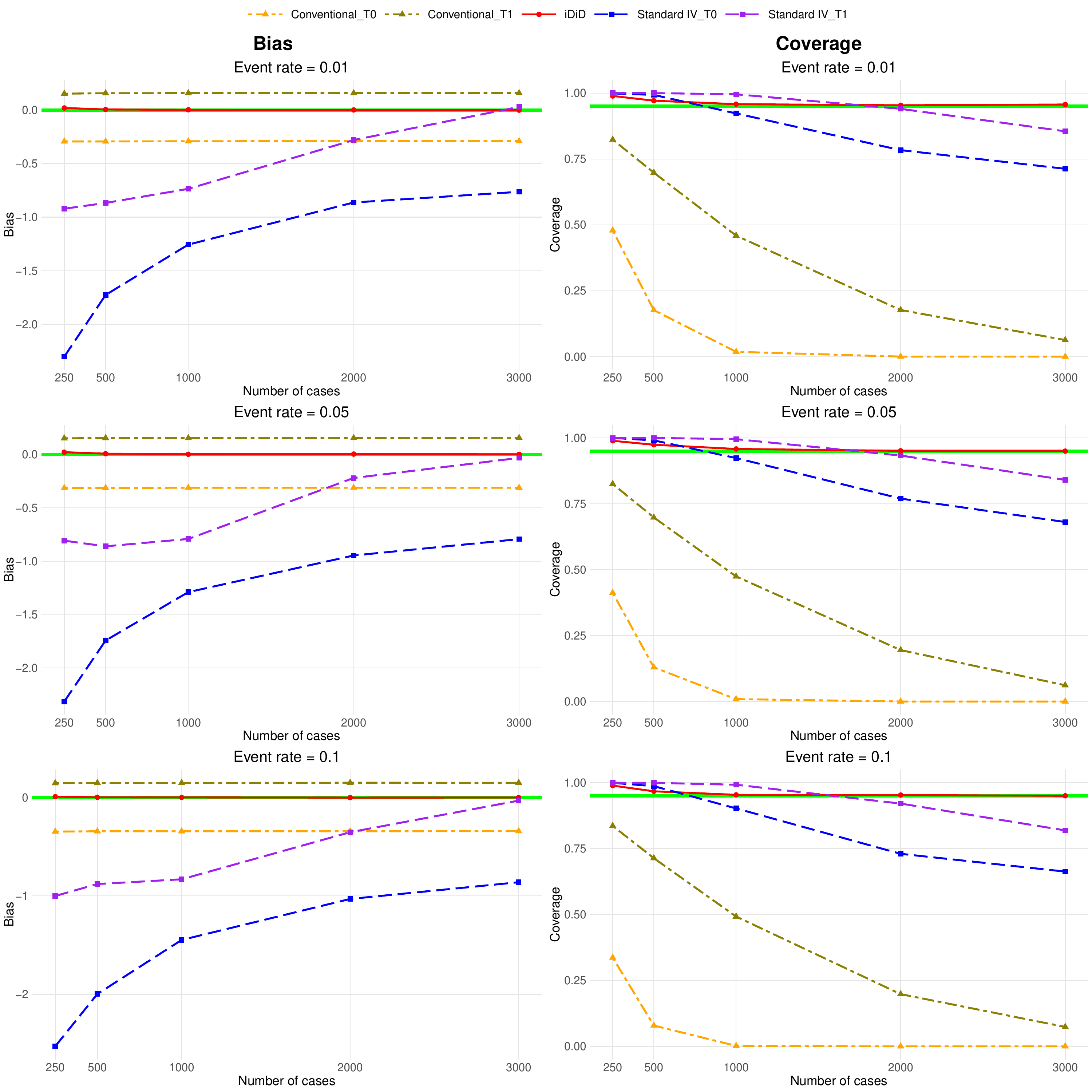}
      \caption{Invalid standard IV}
      \label{fig:invalid_ncross}
    \end{subfigure}
    \caption{Simulation study: Scenario 1 -- Longitudinal data structure. $(a)$: $Z$ is a valid standard IV, $(b)$: $Z$ does not satisfy the standard exclusion restriction assumption due to its time-invariant direct effect on the outcome at each time point.
      Bias of the treatment effect estimate and coverage of the corresponding 95\% confidence interval are evaluated across settings with different event probabilities (1\%, 5\%, 10\%) and sample sizes (250–3000). Three methods are evaluated, including (i) IDiD, (ii) standard IV analysis (using data at either time points), and (iii) conventional case-control analysis that ignores unmeasured confounding (using data at either time points). Green lines indicate the reference values (0 for bias, 0.95 for coverage).}
\end{figure}
\begin{figure}[H]
    \centering
    \begin{subfigure} {0.75\textwidth}
    \centering
      \centering
      \includegraphics[width=1\textwidth]{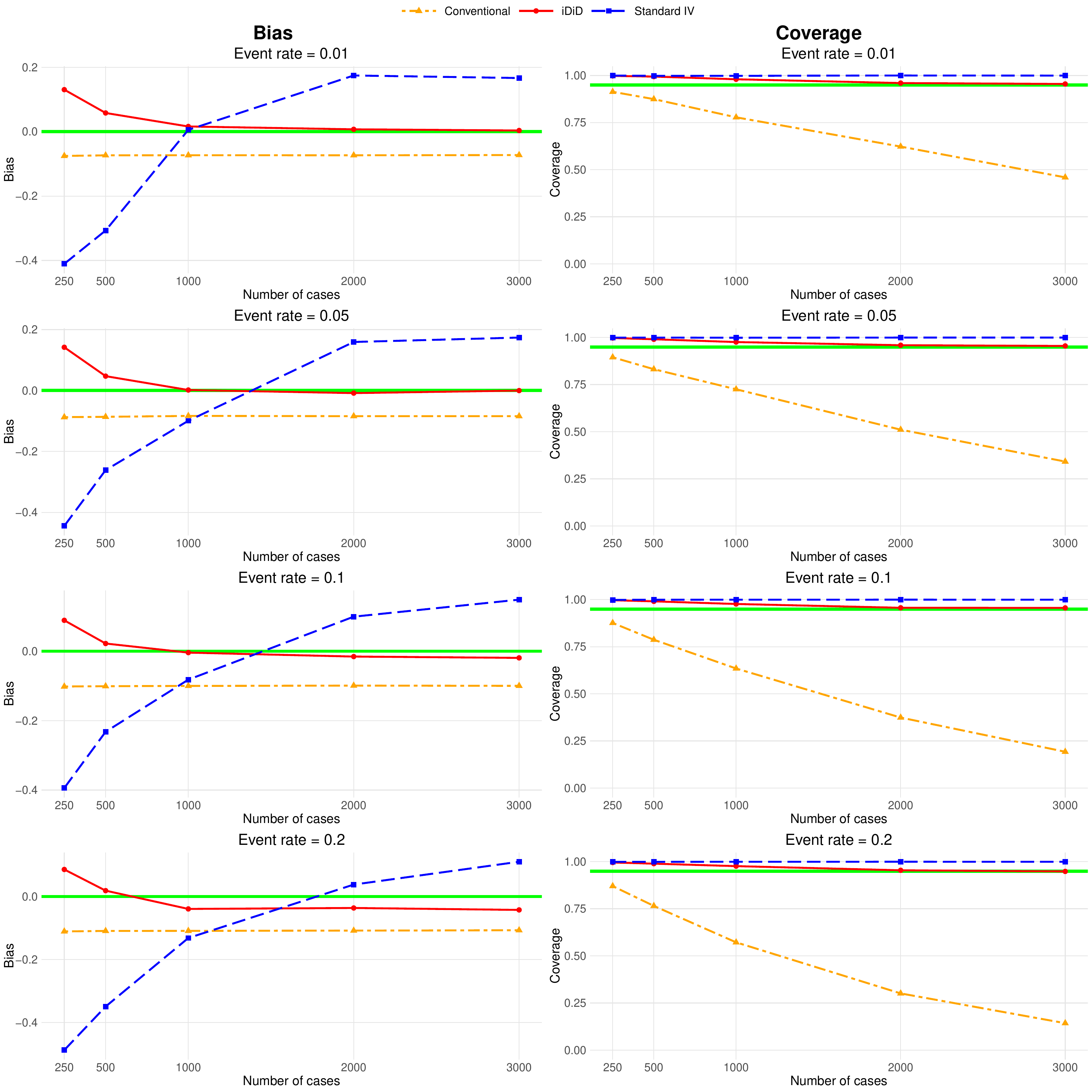}
      \caption{Valid standard IV}
      \label{fig:valid_cross}
    \end{subfigure}
    \begin{subfigure} {0.75\textwidth}
      \centering
      \includegraphics[width=1\textwidth]{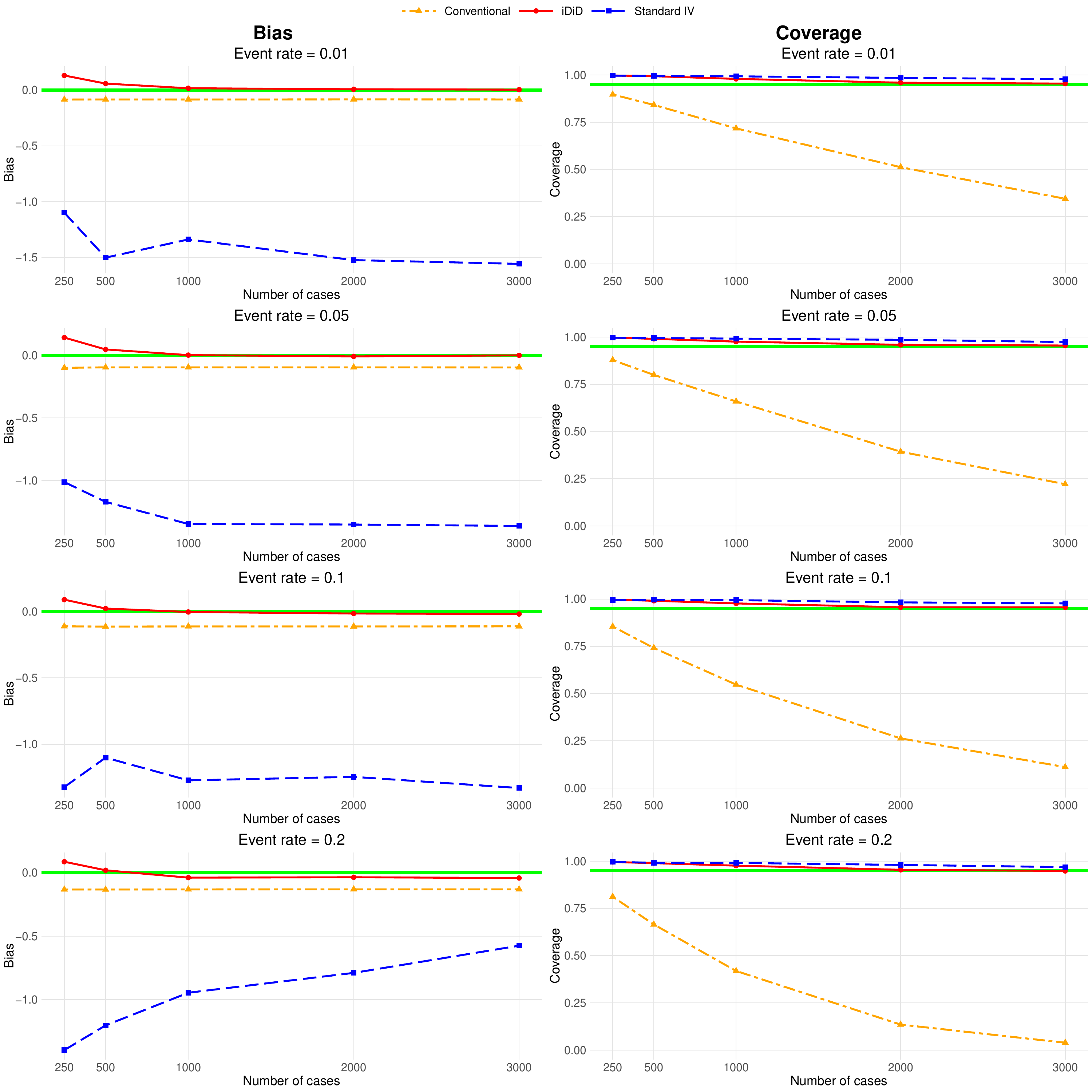}
      \caption{Invalid standard IV}
      \label{fig:invalid_cross}
    \end{subfigure}
    \caption{Simulation study: Scenario 2 -- Repeated cross-sectional data structure. $(a)$: $Z$ is a valid standard IV, $(b)$: $Z$ does not satisfy the standard exclusion restriction assumption due to its time-invariant direct effect on the outcome at each time point.
    Bias of the treatment effect estimate and coverage of the corresponding 95\% confidence interval are evaluated across settings with different event probabilities (1\%, 5\%, 10\%) and sample sizes (250–3000). Three methods are evaluated, including  (i) IDiD, (ii) standard IV analysis (ignoring the temporal aspect), and (iii) conventional case-control analysis. Green lines indicate the reference values (0 for bias, 0.95 for coverage).}
\end{figure}    
\end{landscape}

\end{document}